# Competing magnetic interactions and magnetoresistance anomalies in cubic intermetallic compounds, Gd$_4$RhAl and Tb$_4$RhAl, and enhanced magnetocaloric effect for the Tb case

Ram Kumar,[1,*] Kartik K Iyer,[1] P.L. Paulose[1] and E.V. Sampathkumaran[2,3,**]

[1]*Tata Institute of Fundamental Research, Homi Bhabha Road, Colaba, Mumbai 400005, India*
[2]*Homi Bhabha Centre for Science Education, Tata Institute of Fundamental Research, V. N. Purav Marg, Mankhurd, Mumbai, 400088 India*
[3]*UGC-DAE-Consortium for Scientific Research, Mumbai Centre, BARC Campus, Trombay, Mumbai 400085, India*

Abstract

We report complex magnetic, magnetoresistance (MR) and magnetocaloric properties of Gd$_4$RhAl and Tb$_4$RhAl, forming in the Gd$_4$RhIn-type cubic structure. Though the synthesis of the compounds was reported long ago, to our knowledge, no attempt was made to investigate the properties of these compounds. The present results of ac and dc magnetization, electrical resistivity and heat-capacity measurements down to 1.8 K establish that these compounds undergo antiferromagnetic order initially, followed by complex spin-glass features with decreasing temperature. These characteristic temperatures are: For Gd case, $T_N$= ~46K and $T_G$= ~21 K, and for Tb, ~32 and ~28 K respectively. Additionally, there are field-induced magnetic effects, interestingly leading to non-monotonic variations in MR. There is a significant MR over a wide temperature range above $T_N$, similar to the behavior of magnetocaloric effect (MCE) as measured by isothermal entropy change ($\Delta S$). An intriguing finding we made is that $\Delta S$ at the onset of magnetic order is significantly larger for the Tb compound than that observed for the Gd analogue near its $T_N$. On the basis of this observation in a cubic material, we raise a question whether aspherical nature of the 4f orbital can play a role to enhance MCE under favorable circumstances - a clue that could be useful to find materials for magnetocaloric applications.

*E-mail: ramasharamyadav@gmail.com
**E-mail: sampathev@gmail.com

## I. INTRODUCTION

For the past several decades, considerable research activities have been focused to understand various phenomena associated with the tendency of 4f electrons to exhibit delocalization, particularly at the beginning of the rare-earth (R) series. While these activities led to path-breaking discoveries in condensed matter physics, generally speaking, the rare-earths with localized 4f electron have attracted less attention till recently. The need to focus on such 'normal' rare-earth materials was stressed by one of us [1] more than two decades ago, pointing out the existence of unusual transport anomalies [2] in the paramagnetic state demanding new concepts in magnetism. These anomalies triggered new theoretical ideas, however, in recent years only [3]. Even in the magnetically ordered state, huge unconventional Hall anomalies in an intermediate field range across metamagnetic transitions were reported two decades ago in $Gd_2PdSi_3$ [4], though such Hall anomalies subsequently in other materials containing transition metal ions led to the proposal of Topological Hall Effect (THE) due to magnetic skyrmions [5]. Needless to mention that the magnetic skyrmions - one of the most exciting topics of research in magnetism - are vortex-like nanometric spin-textures, bearing a potential for spintronic and next generation information storage devices. The transport anomalies briefed above as well as demonstration of the existence of competing exchange interactions triggered other groups [6] in recent years to look for magnetic skyrmions successfully not only in $Gd_2PdSi_3$, but also in $Gd_3Ru_4Al_{12}$ [Ref. 7] as reviewed in Ref. 8. In this respect, $GdRu_2Si_2$ is also of great interest [9]. Incidentally, GdPtBi is considered to be a Weyl semimetal [10]. Needless to state that the pioneering work on magnetocaloric effect (MCE) was on a Gd compound [11], which triggered a lot of activity in this field. Clearly, the searches for new materials containing well-localized 4f orbital, particularly to identify those showing competing exchange interactions and anomalous transport behavior, may yield results that would aid future discoveries of novel concepts, even after decades of such reports, judged by the past experience, as briefed above. In this respect, for Gd cases, even the interference from the crystal-field effects can be ignored.

Keeping this in mind, we focus our attention on R-Rh-Al ternary family with 4:1:1 stoichiometry which has not been paid any attention in the literature [12]. We have performed bulk measurements and the major findings are: (i) Competition between ferromagnetic and antiferromagnetic correlations as a function of temperature ($T$) and magnetic field ($H$), also leading to reentrant spin-glass anomalies with multiple relaxation rates as well as magnetoresistance anomalies (MR). (ii) There are considerable efforts in the current literature to find materials for magnetic refrigeration in various temperature ranges, particularly near room temperature [13-15]. It is therefore important to explore various factors which can lead to large MCE properties. We report here that the Gd compound shows a relatively weaker MCE with respect to Tb case. This intriguing finding prompts us to raise a question whether single-ion anisotropy of the orbital responsible for magnetism is a key to enhance MCE under favorable circumstances. Cubic structure of these materials enables us to infer such a possible scenario, as it is difficult to separate out crystalline anisotropic effects in non-cubic structures.

Attempts to study these Rh compounds were primarily motivated by the observation [16-18] of interesting magnetic and transport anomalies in Pt analogues, $R_4PtAl$ (R= Gd, Tb and Dy), the synthesis of which was reported by Engelhart and Janka [19] a few years ago. Many of these compounds with 4:1:1 stoichiometry have been known [12, 19, 20, 21] to form in the $Gd_4RhIn$-type cubic structure ($F\bar{4}3m$) and it is surprising that this R-Rh-Al family was not subjected to detailed magnetic and transport investigations, following the synthetic report by Tappe et al [12] several years ago. This cubic structure is characterized by interesting crystallographic features, which we already presented in our earlier publication on Pt analogue [17]. The most relevant feature from magnetism angle is that there are three crystallographically inequivalent sites for R. Such multiple sites for R often in general have been known to lead to magnetic and transport anomalies, as introduced in Ref. 16, the most popular one (in the field of MCE) being $Gd_5Si_2Ge_2$ with five crystallographically inequivalent positions for Gd [11]. In the 4-1-1 family, the $Gd_4PtAl$ exhibits antiferromagnetism ($T_N$= 64 K) to spin-glass transition ($T_G$= 20K) [16] and $Tb_4PtAl$ [17] shows spin-glass features at the onset of antiferromagnetic order (at 50 K), while $Dy_4PtAl$ is characterized by ferromagnetic order at 32 K followed by spin-glass anomalies around 20 K [18]. There are evidences for a magnetic-field-induced first-order transitions as well. A noteworthy observation we made in this Pt family is that the value of $\Delta S$ at the onset of magnetic order in the case of Dy is significantly larger compared to that for Gd compound. Besides, analogous Cd-based compounds [22-26] have been reported to exhibit large MCE. Some notable examples are: For $Er_4PdMg$: $\Delta S$= 22.5 J/kg-K at its Curie temperature ($T_C$) of 16 K [Ref. 23]; $Eu_4PdMg$: 7.2 J/kg-K at its $T_C$ of 150 K, and a table-like curve over a wide temperature ($T$) range [22], a feature required for applications. Motivated by these observations, we considered it worthwhile to probe magnetic and magnetocaloric behavior of the Rh members as well.

## II. EXPERIMENTAL DETAILS

The polycrystalline samples were prepared by arc-melting together stoichiometric amounts of high purity constituent elements [R: >99.9%; Rh: >99.99%; Al: >99.999%] in an atmosphere of argon. Powder x-ray diffractions (Cu $K_\alpha$) and Rietveld analysis, shown in figure 1, confirm that the samples form in the $Gd_4RhIn$-type cubic structure, without any impurity within the detection limit (<2%) of x-ray diffraction method. Scanning electron microscopic technique was also employed to verify composition homogeneity. $T$-dependent (down to 1.8 K) dc magnetization ($M$) as well as isothermal $M$ measurements at selected temperatures were carried out employing a commercial (Quantum Design) vibrating sample magnetometer; a commercial (Quantum Design) SQUID magnetometer was employed to measure ac susceptibility ($\chi$) with some frequencies (with an ac field of 1 Oe). Heat-capacity ($C$) by relaxation method and dc electrical resistivity ($\rho$) studies by the four-probe method were performed with the help of a commercial Physical Properties Measurements System (PPMS, Quantum Design). Unless stated explicitly, all the measurements were done for the zero-field-cooled (ZFC) state of the specimens.

## III. RESULTS AND DISCUSSION
### A. Magnetic anomalies in $Gd_4RhAl$

Figure 2a shows dc $\chi$ behavior, measured in a magnetic-field of 5 kOe, as a function of $T$ for $Gd_4RhAl$. Inverse $\chi$ follows Curie-Weiss behavior down to about 75 K. The effective moment ($\mu_{eff}$) obtained from this linear region is ~8.1$\mu_B$/Gd, which is in good agreement with that expected for free $Gd^{3+}$ion (7.94 $\mu_{eff}$). This suggests that there is no magnetic moment on Rh. The value of paramagnetic Curie temperature ($\theta_p$) is ~21K and the positive sign implies dominant ferromagnetic correlations. However, as the results reveal below, the magnetic ordering that sets in is of an antiferromagnetic type. As the temperature is lowered, in the plot of $\chi(T)$ for $H$= 5 kOe, following a monotonic increase, there is a broad shoulder in the range 42-47 K, as though the magnetic order sets in around this temperature; subsequently, there is a well-defined peak at ~21K, as though there is another magnetic transition. The curves obtained in a field of 100 Oe for ZFC and field-cooled (FC) conditions of the specimen show a bifurcation near 45 K, that is, at the onset of magnetic order (figure 2b); the fall below the peak for FC curve is relatively less compared to that in ZFC curve, possibly indicating spin-glass-type of ground state [27]. We have also measured isothermal remnant magnetization ($M_{IRM}$) at selected temperatures across this 21 K transition, e.g., at 5, and 15 K. For this purpose, we cooled the sample to these desired temperatures in the absence of an external magnetic field, switched on a field of 5 kOe, kept for 5 mins and then switched off. We then measured magnetization ($M_{IRM}$) as a function of time ($t$). We find that $M_{IRM}$ decays with $t$ rather slowly below 21 K, as shown in figure 2c. We also took this data at 40 K (just below the magnetic ordering temperature) and no such decay of $M_{IRM}$ was found with the values remaining essentially constant with time. This observation supports spin-glass freezing below 21 K. (Therefore, the divergence of ZFC-FC curves near 45 K is not due to spin-glass freezing and may be associated with domain wall effects). Our attempts to fit this relaxation curve to a single exponential form of the type $M_{IRM}= M_{IRM}(0) + Ae^{-(t/\tau)}$ (or a stretched form), where A is a constant and τ is the relaxation time, is not satisfactory; instead, a better fitting is obtained with three exponential decay terms. The three τ values thus obtained are rather large - about 100, 500 and 3200 s - with a marginal temperature dependence, indicating that *this magnetic phase is a complex multi-relaxation cluster-glass state*, possibly associated with three sites for R.

In order to substantiate that there is no ferromagnetic ordering, we show in figure 2d, the isothermal $M$ curves (virgin) at 2, 10 and 30 K in the units of magnetic moment per formula unit (f.u.). We noted that there is a weakly hysteretic $M(H)$ for 2 K only, that too in the low field-range, but no hysteresis could be resolved for 10 and 30 K. But the fact remains that there is no evidence for saturation even at fields as high as 130 kOe and the curvature persists. It appears that there are subtle changes in the slope of $M(H)$ at low fields at 2 K, as inferred by peaks in the derivative plot near 960 Oe and 7.35 kOe (see inset of figure 2d), though these are smeared for 10 and 30 K. [This observation corroborates well with that seen in isothermal MR data presented later in this article]. Studies on single crystals may be required to confirm such a low-field feature. It appears that there is a change of slope of $M(H)$ curve even at a very high field (~85 kOe). These results, apart from suggesting the existence of multiple field-induced magnetic transitions, establish that ferromagnetism is absent in this material in zero field, and antiferromagnetism persists even at 2 K. We therefore infer that antiferromagnetism may coexist with spin-glass freezing below 21 K, and we can not rule out the possibility that these different types of order arise from crystallographically inequivalent Gd ions. Dominance of antiferromagnetism in the magnetically ordered state is interesting despite that the sign of $\theta_p$ is positive which is indicative of tendency towards ferromagnetism. The fact that the value of Néel temperature ($T_N$) is more than that of $\theta_p$ is also in favor of dominant antiferromagnetic interaction. Thus, there is presumably a magnetic frustration due to a competition between antiferromagnetic and ferromagnetic correlations.

To further render support to the above-mentioned magnetic behavior, we show $C(T)$ in figure 3. Heat-capacity curve exhibits an upturn below 47 K and a peak at 44.7 K. A peak establishes the existence of a long-range magnetic order of a well-defined magnetic structure near 46 K. However, the jump in $C(T)$ at the onset of magnetic order (< 5 J/Gd mol-K) is far less than that expected (20.15J/Gd mol-K, Refs. 28, 29) for equal-moment or ferromagnetic ordering; such a strong reduction implies a complex antiferromagnetic structure, e.g., an amplitude modulated type. There is no anomaly near 21 K except a very weak and broad shoulder spreading over a wide $T$-range of about 20-30 K (figure 3), which is consistent with the fact that the 21K-transition is of a glassy character. We have noted that the value of the linear term extrapolated to absolute zero temperature, inferred from the plot of $C/T$ versus $T^2$ below 4 K, is rather large (about 300 mJ/mol $K^2$). Such a large value can not be attributed to heavy-fermion behavior for such heavy rare-earth members with strictly localized 4f orbital and therefore has to arise from cluster-glass behavior [see, for instance, Ref. 30]. In fact, *C* varies quadratically over a wide temperature (figure 3, top inset), showing neither $T^{1.5}$ form nor $T^3$ form expected for ferromagnets/spin-glass and antiferromagnets respectively [see, for instance, 30-32]. These results establish that there is a complex reentrant magnetism in this compound. In the presence of external fields (measured up to 50 kOe), the peak is quite intact (figure 3, right inset) without any noticeable change in the functional form at lower temperatures. A careful look at the curves suggests that the values of *C* in 50 kOe are lower with respect to those in zero-field at $T<<40$ K and the opposite situation is observed for $T>40$ K up to $T_N$. The exact reason for this crossover is not clear to us at present. We believe that such a complex situation could be due to the combined effect of magnetic field on the $2J+1$ degeneracy of Gd 4f [29] and a gradual change in the magnetic character as the temperature is lowered below 40 K. We will return to further relevance of in-field data to derive MCE later in this article.

We now discuss the results of ac χ measurements. The real part (χ') measured in the absence of an external magnetic field, shown in figure 4, exhibits a pronounced peak at about 21 K and the imaginary part (χ'') also exhibits a peak; it appears that there is a weak frequency (υ) dependence with the amplitude decreasing with increasing υ, as shown in the inset of figure 4a, as the υ is varied from 1.3 Hz to 1.333 kHz. These peaks vanish for an application of a small dc field of 10 kOe. These findings are consistent with what is expected for spin-glass freezing [27]. However, in the region, 42-47 K, in the zero-field curves, there is only a very weak υ-independent shoulder in χ' and no such feature could be resolved in χ''. This finding supports that the magnetic ordering setting in around 46 K, as revealed by *C(T)* peak, is not of a spin-glass-type.

In figure 5a, we present the results of dc electrical resistivity measurements as a function of *T* in the presence of a few external magnetic fields. In zero field, following positive temperature coefficient of ρ in the high temperature range (see figure 5a, inset), there is a sudden drop at $T_N$ due to the loss of spin-disorder contribution. There is a shoulder near 21 K as the material enters spin-glass phase. Comparing the zero-field and in-field curves (see the mainframe of figure 5a), over a wide temperature range in the paramagnetic state (~46 to ~90 K), the values are reduced with *H*; in other words, the magnetoresistance, defined as MR= {ρ(*H*)-ρ(0)}/ρ(0), is negative. Such a negative MR over twice of $T_N$ has been reported by us in the past in many systems suggesting interesting magnetic precursor effects [1, 2]. In the magnetically ordered state, though MR is negative for 50 kOe, there is a sign crossover for 10 and 30 kOe. In order to address this issue further, we have taken isothermal data at several temperatures. As shown in figure 5b, for 1.8 and 10 K, there is indeed a sign-crossover near 40 kOe, with the values of MR being positive initially. The curves for these temperatures reveal that there is a competition between positive contribution (dominating at low fields) and negative contribution due to the tendency of the magnetic-field to align the magnetic moments. A careful look at the curves at higher fields suggest that there is another slope change near 100 kOe, which appears to be consistent with the high-field feature seen in the corresponding *M(H)* curve. With increasing temperature, say at 25 and 40 K, antiferromagnetism is expected to contribute at low-fields, but the negative sign at all fields, instead of positive sign implies that magnetic Brillouin-zone gaps (antiferromagnetic energy-gap) form in this magnetic phase. Finally, a careful look at the hysteresis curve at 1.8 K, shown in figure 5b (inset) in the range -50 to + 50 kOe is quite revealing. That is, the virgin curve lies outside the hysteresis loop. This is a characteristic feature of disorder-broadened first-order magnetic transition [33]. This implies that the zero-field state contains an antiferromagnetic phase, as, such a behavior of MR loop is not expected for conventional metallic contribution from conduction electrons alone. The positive sign of MR implies that the antiferromagnetic structure is different from the one in the range 21-46 K. Clearly, this phase undergoes a spin reorientation for an application of magnetic fields (close to 10 kOe inferred from *M(H)* curves) and the high-field state thus obtained is characterized by negative MR. The hysteresis gets negligible with increasing temperature (e.g., at 10 K).

All these results establish that this compound undergoes an antiferromagnetic order near 46 K, with a complex spin-glass phase developing below 21 K; an antiferromagnetic component also appears to coexist below 21 K.

### B. Magnetic anomalies in Tb$_4$RhAl

The results of $\chi(T)$ measurements in 100 Oe and 5 kOe are plotted in figure 6a. The Curie-Weiss behavior is seen above 50 K and the value of $\mu_{eff}$ (~9.97$\mu_B$/Tb) is very close to that known for free Tb$^{3+}$ ion (9.72 $\mu_B$). The value of $\theta_p$ is ~ -3 K and the negative sign unlike in the Gd case implies dominant antiferromagnetic intersite interaction. However, though the magnitude of $\theta_p$ is small, as the *T* is lowered, there is a sudden increase in the slope at 32 K (see the curves in 100 Oe, figure 6b). This suggests that the magnetic ordering sets in at this higher temperature. Thus, this characteristic temperature bears no correlation with $\theta_p$, pointing to the existence of competing exchanging interactions. As evidenced below, this transition is of an antiferromagnetic type, intersected by another transition at 27 K. That is, at 27 K, there is a peak in 100 Oe ZFC curve and a shoulder in 5 kOe. ZFC-FC curves measured in 100 Oe separate below 27 K; but the values of $\chi$ for the FC condition keeps increasing with decreasing *T* down to 4 K which is typical of cluster spin-glasses [30, 34]. $M_{IRM}$ in zero-field measured at different temperatures below 27 K as per the protocol briefed above decays gradually with time, as shown in figure 6c for 5 and 25 K. As in the case of Gd, it appears that the spin-glass state is characterized by three relaxation times (around 100, 450 and 3500 s).

Incidentally, there is a peak at 8 K in the data measured with 5 kOe and the shape of $\chi$ is seemingly different from the ZFC curve obtained in 100 Oe; this is an artifact of magnetic-field-induced changes in magnetization, as shown in figure 7 in the range -40 kOe to 40 kOe for 1.8, 5, and 10 K. In the virgin *M(H)* curve of 1.8 K, there is an increase in the slope of *M(H)* near 10 kOe for 1.8 K, 3 and 12 kOe for 5 K, and 1 and 8 kOe for 10 K. This feature at 1.8 K interestingly becomes discontinuous after a field cycling and such an observation has been reported by us in the past for some other antiferromagnetic rare-earth based intermetallics [35], but the explanation for such a field-cycling effect is not known yet. It is obvious from these figures that the virgin curve lies outside the hysteresis loop, thereby suggesting that these transitions are disorder-broadened first-order transitions. Thus, the zero-field spin-glass state appears to be a complex one with possible coexistence with an antiferromagnetic segment, also responsible for the above-mentioned field-induced transitions. In order to see how *M(H)* behaves between 28 and 32 K, we show the *M(H)* curve at 28.5 K also in the 10K-panel. We see an increase in slope around 20 kOe (apart from a sharp one at low-fields) – a transition field which is higher than that noted for 10 K - and the curve is not hysteretic. These demonstrate that the antiferromagnetic part above and below 28 K may be different and the transition at 32 K must be of an antiferromagnetic type considering the existence of 20kOe spin-reorientation feature. Finally, we would like to mention that *M* keeps increasing with *H* when measured up to 140 kOe without any evidence for saturation (shown for 1.8 K in the inset of figure 7), however with an observable slope change around 120 kOe (which we noted to persist in the 5, and 10 K data, not shown). This might indicate that ferromagnetism is not attained even at very high fields.

In figure 8, we show the heat-capacity behavior. In zero-field, there is a distinct λ-anomaly at 32 K, establishing the onset of long-range magnetic order. We are not able to resolve any other peak, though there is a broad shoulder at a few degrees below 32 K (in the vicinity 28 K), which endorses the inference for another magnetic transition from the χ(*T*) curves presented above. As in the case of Gd compound, if one looks at the linear regime (that is, below 6 K) in the plot of *C/T* versus $T^2$, a large value (of about 390 mJ/mol $K^2$) for the linear term, characteristic of cluster spin-glasses, is obtained. *C* varies quadratically over a wider temperature range (till about 10 K, see the inset of figure 8). Finally, for an application of magnetic field of 30 kOe, the peak is shifted to a marginally lower temperature, indicative of antiferromagnetism for 32 K transition. For a further higher field of 50 kOe, the peak gets smeared, with the curve shifting towards a marginally higher temperature, and this reflects the tendency towards ferromagnetic alignment (though total ferromagnetic alignment is absent as inferred from magnetization data).

We present the results of ac χ measurements with different frequencies in figure 9. There is a peak (in zero field) in both χ' and χ" at 28.2 K, but not at the temperature (32 K) where the material enters magnetically ordered state. These peaks in zero-field are frequency-dependent, though weak (see the inset of figure 9) as known for canonical spin glasses. In a field of 5 kOe, these peaks vanish, which is consistent with the proposal of spin-glass freezing at 28.2 K; however, a weak hump is transparent in χ' at 32 K; this must be attributed to non-glassy ordering (present in zero-field as well), as χ" is featureless in 5 kOe.

Now looking at electrical transport behavior of ρ (see figure 10a), positive *d*ρ/*dT* in the paramagnetic state is observed, as expected for metallic systems. A kink in ρ appears at the onset of antiferromagnetic order, but no broad feature is apparent at the spin-glass transition, presumably because it is very close to $T_N$. With the application of external fields (10, 30, and 50 kOe), these kinks get smeared, though the drop due to magnetic ordering persists with the lowering of temperature. As in the case of Gd compound, negative MR is seen at temperatures well above $T_N$. In order to see the features due to magnetic-field induced transitions inferred from *M*(*H*) curves, we show in figure 10b the virgin isothermal MR(*H*) curves for 1.8, 5, 10 and 28.5 K, apart from showing the curve for one temperature (40 K) in the paramagnetic state. In the paramagnetic state, quadratic field-dependence, expected for paramagnets, is seen. Barring a small value in the positive quadrant below about 10 kOe for 1.8 K, the curves lie in the negative quadrant. The magnitude of MR in the positive quadrant is so small that it is very difficult to infer whether it is due to the influence of magnetic field on the conduction electrons or whether it is due to antiferromagnetic part. In any case, the sign cross-over is visible for 1.8 K. Though the features due to two closely spaced magnetic-field-induced transitions at low fields, inferred from the *M(H)* data, could not be clearly resolved in the curves for 1.8, 5 and 10 K, one could see profound changes in MR in the range 0 to 20 kOe at different fields depending on the temperature of measurement. For 28.5 K, the existence of 20kOe-transition, inferred from *M*(*H*) curves, finds support from the sudden change in the slope. It is to be noted that there is a broad hump in the MR(*H*) curves around 100 kOe for 1.8, 5 and 10 K, indicative of another magnetic anomaly at this field, supporting an inference from *M*(*H*) data. Finally, we show in figure 10c the isothermal MR loop behavior (+120 to -120 kOe) at 1.8 K to get a better insight into the transitions. The loop is hysteretic and it is distinctly clear that the virgin curve lies outside the envelope curve for initial applications of *H,* thereby providing compelling evidence for the first-order nature of the field-induced transitions at low fields. It may be remarked that the absolute magnitude of MR, say at 50 kOe, is much larger than that observed for the Gd analogue at 1.8 K. Thus, this Tb sample is relatively more magnetoresistive.

## C. Comparison of magnetocaloric effect

We now compare and contrast magnetocaloric behavior of these two compounds. For this purpose, one can obtain $\Delta S$ by two different methods, one by using the $T$-dependent heat-capacity curves in the presence of desired magnetic fields (shown in figures 3 and 8) and the other using from isothermal magnetization curves measured at close intervals of temperature (typically 2 K) using Maxwell equation. Since magnetization curves for selected temperatures are shown in figures 2d and 7, we have not shown the curves measured at closer interval of temperatures. We plot -$\Delta S$ values for a few final fields in figures 11a and 11b obtained from above-mentioned two methods. As expected, the curves obtained by both the methods are very close to each other. Looking at the curves in figure 11a, for the Gd compound, the sign of -$\Delta S$ is negative over a wide $T$-range above 1.8 K, say till about 22 K for $H$= 10 kOe, which implies dominance of antiferromagnetic component [14] over spin-glass phase to magnetocaloric effect; at higher temperatures, the values lie in the positive quadrant suggesting dominance of ferromagnetic correlations in the presence of such magnitudes of external fields. Similar behavior is seen for Tb$_4$RhAl (figure 11b). The values exhibit a maximum near respective magnetic ordering temperatures. However, an intriguing finding is that the peak value for a given value of $H$ is 2 to 3 times larger for the Tb compound, compared to that for Gd compound. For instance, for $H$= 50 kOe, the maximum value of -$\Delta S$ is about 2.3 J/kg-K for Gd case, whereas the value is about 6.5 J/kg-K for Tb. Our preliminary studies on Ho and Er members of the present series also reveal much higher magnitudes of $\Delta S$ compared to that for Gd analogue. In the Pt series also, we made similar observations. For Gd$_4$PtAl, Tb$_4$PtAl and Dy$_4$PtAl, the peak values of -$\Delta S$ are ~6 J/kg-K, 6 J/kg-K, and 13 J/kg-K for a change of $H$ from 0 to 50 kOe; in this family, Dy compound is characterized by a relatively higher peak value, and not Gd. While it is known [see, for instance, 36-39] that crystalline anisotropy can favor larger MCE for some directions, possible role of aspherical nature of the 4f orbital (single-ion anisotropy of the relevant orbital) to enhance MCE in some families of rare-earths was not explored in the past literature. In this connection, we draw the attention of the readers to the trend within some other families. An inspection of Tables 3 (for RFeAl) and 6 (for RCo$_2$B$_2$C and RNiBC) in Ref. 13 suggests that maximum MCE is not for the corresponding Gd compound within respective rare-earth families, and therefore the speculation made here (that is, aspherical nature of the orbital responsible for magnetism can under favorable circumstances promote better MCE) is a food for thought for some situations, deviating from conventional wisdom of attributing to subtle differences in magnetic behavior among the compounds in a series. Finally, a visual inspection of the curves in figure 11 suggests that the Relative Cooling Power (RCP), which is defined as the product of the full-width at half maximum in the plot of $\Delta S$ versus temperature and maximum entropy change must be relatively larger for the Tb case compared to the Gd compound. Though it is not easy to determine exact values of RCP due to the asymmetric nature of the curves (particularly for the Gd case) and sign cross-overs, an estimate can be made for a comparative purpose by Gaussian fitting to the positive quadrant of the curves. These values turn out to be about 110 and 340 J/kg for $\Delta H$= 50 kOe for Gd and Tb compound respectively – clearly higher for the Tb case. These values may be compared with RCP of well-known MCE materials for an applied field typically of 50 kOe, e.g., like Gd$_5$Si$_2$Ge$_2$ [11], MnAs [40], Gd$_6$Co$_{1.67}$Si$_2$ [41], Nd$_2$NiMnO$_6$ [42], and recently MnPdGa (with a smaller $\Delta S$ value of 3.5 J/kg-K, [43]) for which RCP= 278, 390, 310, 300, and 225 J/kg respectively. In the temperature range of interest for the present compounds (20-50 K), as inferred from the Tables in Ref. 13, HoFeAl, TbCo$_{0.5}$Al$_{1.5}$, and DyCo$_{0.5}$Al$_{1.5}$, are characterized by RCP values in the range 310 – 370 J/kg and higher values (<527 J/kg) are reported for a few (GdCo$_2$B$_2$, Ho$_2$Cu$_2$In and Ho$_4$PtMg) only; many other compounds show much lower RCP values. Finally, the extension of the tail of $\Delta S$ to a wider temperature range above respective $T_N$ is similar to the MR behavior; such a correlation between MR and MCE was reported in the past as well [44]. It is therefore tempting to propose that MR measurements serve as another route to identify large MCE materials.

We show the plots of $\Delta S$ as a function of $H$ for different temperatures around respective magnetic onset temperatures for both the compounds in figure 12 and fitted to the power law $H^n$. It is found that there is a deviation from the mean-field value of 2/3 for $n$ predicted long ago [45]. The values of the exponent exceed even 1. Such a deviation has been reported in the literature [46] for some materials and a quantitative criterion has also been evolved recently [47] according to which second-order transitions are characterized by a large spread in the value of $n$. The present results are consistent with second-order magnetic transition.

## IV. SUMMARY

This is the first detailed study of magnetic, magnetotransport and magnetocaloric behavior of two compounds in the R-Rh-Al ternary family with 4:1:1 stoichiometry, yielding intriguing results. There are distinct evidences for at least two temperature-induced magnetic transitions – viz., paramagnetism to antiferromagnetism to a complex spin-glass (as indicated, for instance, by multiple relaxation times). Besides, an antiferromagnetic component seems to coexist with the spin-glass phase. Thus, the magnetism in these materials is characterized by competing exchange interactions among the three magnetic sites of R. Competition between antiferromagnetic and ferromagnetic interactions are also inferred from (i) the paramagnetic Curie temperature values (with respect to those of $T_N$), (ii) the features attributable to magnetic-field induced spin-reorientation transitions at low temperatures as inferred from isothermal magnetization and MR curves at 1.8 K, and (iii) the sign changes in $\Delta S$ as a function of $T$. In view of these, it is of interest to explore whether magnetic skyrmions, particularly in the intermediate field ranges, form.

With respect to the behavior of corresponding Pt compounds [16-18], there are some subtle differences (e.g., positive MR in the magnetically ordered state due to dominant antiferromagnetic component and multiple spin-glass features for Tb$_4$PtAl – but not seen for Tb$_4$RhAl). The quadratic $T$-dependence of heat-capacity in the present compounds unlike in Pt family also reveals complexity of magnetism.

This work adds to the data base of magnetic materials in which more than two sites are present for the same magnetic ion and hence would serve as model systems for theoretical formulation for magnetic frustration due to multiple magnetic sites, arising from complicated Fermi surface [see, for instance, Ref. 48, for the idea of 'topological spin crystals due to itinerant magnetic frustration' which includes magnetic skyrmions].

A conceptually interesting speculation made here based on the comparison of peak values of $\Delta S$ as well as estimated RCP of the Gd and Tb compounds is that the anisotropic nature of the orbital responsible for magnetism seems to favor better magnetocaloric effect. Usually, whenever the compound under study is non-cubic, observed enhancement of MCE for a specific orientation of single crystals is attributed to the role of crystalline anisotropy [36-39]. Since the present compounds are cubic, this family of materials has provided a unique opportunity to propose the possible role of asphericity of the orbital concerned on MCE. We believe that this factor could play a role in other noncubic materials as well. Needless to add that, as established long ago [49-51], the sign of the crystal-field terms for the Hamiltonian (that is, the anisotropy of the crystal-field-split 4f ground state) has been shown to influence the magnetic ordering temperature, and recently also the magnetodielectric effect [52]. This work provides an indication for such effects on another phenomenon of application potential, while we welcome future efforts to test the validity of this speculation. If found valid, this would be a clue to search for new materials for magnetocaloric applications, particularly at room temperature. It is of interest to develop theories of MCE with this factor in mind.

## ACKNOWLEDGMENTS

One of the authors (E.V.S) would like to acknowledge the initial support of Science and Research Engineering Board for awarding J C Bose Fellowship (Sanction Order No. SB/S2/JCB-23/2007), and the subsequent support at the final stages by Department of Atomic Energy, Government of India, by awarding Raja Ramanna Fellowship.

**Figure captions**

Figure 1:
X-ray diffraction patterns (Cu K$_\alpha$) of Gd$_4$RhAl and Tb$_4$RhAl at 300 K. The continuous lines through the data points are obtained by Rietveld fitting. Vertical bars denote positions expected for diffraction lines and the difference between experimental and fitted line is also shown. Rietveld fitted parameters are also given.

Figure 2:
Magnetic susceptibility ($\chi$) and inverse susceptibility obtained in a field of (a) 5 kOe by zero-field-cooling (ZFC) and (b) 100 Oe by ZFC as well as field-cooling (FC) for Gd$_4$RhAl. The lines through the data points serve as guides in all curves, except in inverse $\chi$ plot in which case the line is obtained by Curie-Weiss fitting.   (c) Profile of isothermal remnant magnetization as a function of time obtained as described in the text for 5 and 15 K. (d) Isothermal magnetization (virgin) curves as a function of magnetic field at 2, 10 and 30 K; the derivative plot is shown for 2 K.

Figure 3:
Heat-capacity (*C*) as a function of temperature (1.8 – 80 K) for Gd$_4$RhAl in the absence of magnetic field. Top inset shows the plot of *C* vs $T^2$ (below 20 K). Bottom inset shows the curves in the region 35 to 50 K in 0 and 50 kOe). Since the curves for 10 and 30 kOe almost overlap with the zero-field curves, these are not shown for the sake of clarity. The lines through the data points serve as guides to the eyes.

Figure 4:
Real ($\chi'$) and imaginary ($\chi''$) parts of ac susceptibility as a function of temperature (1.8-60 K) for Gd$_4$RhAl in zero field (and in 10 kOe for χ'), measured with different frequencies. Inset shows χ' plots in an expanded form around 24 K to show the frequency dependence. In χ" plot, the (zero-field) curves for 1.3 and 133 Hz only are shown for the sake of clarity. A vertical arrow in the zero-field χ' plots marks the feature at the magnetic transition around 47 K.

Figure 5:
(a) Electrical resistivity as a function of temperature (1.8-100 K) for Gd$_4$RhAl in the presence of different external magnetic fields. Inset shows the curve up to 300 K, obtained in the absence of external field. (b) Magnetoresistance as a function of magnetic field (0 – 130 kOe, virgin curve only) for 1.8, 10, 25 and 40 K. Inset shows the MR versus $H$ behavior at 1.8 K for a variation of the field, 0 to 50 kOe to -50 kOe to 50 kOe to 0, and the arrows mark the way the field is varied.

Figure 6:
Magnetic susceptibility (χ) and inverse susceptibility obtained in a field of (a) 5 kOe by zero-field-cooling (ZFC) and (b) 100 Oe by ZFC as well as field-cooling (FC) for Tb$_4$RhAl. The lines through the data points serve as guides in all curves, except in inverse χ plot in which case the line is obtained by Curie-Weiss fitting. (c) Isothermal remnant magnetization as a function of time obtained as described in the text for 5 and 25 K.

Figure 7:
Isothermal magnetization hysteresis loops at 1.8, 5 and 10 K for Tb$_4$RhAl and the arrows are drawn to show the way the field is being varied. Inset in 1.8K shows the profile up to 130 kOe in the positive quadrant and the one in 10K panel shows the virgin curve for 28.5 K up to 120 kOe.

Figure 8:
Heat-capacity as a function of temperature (1.8 – 60 K) in 0, 30 and 50 kOe for Tb$_4$RhAl. Inset shows the plot of $C$ versus $T^2$ for the zero-field data in the low temperature region.

Figure 9:
Real (χ') and imaginary (χ") parts of ac susceptibility as a function of temperature (1.8-40 K) for Tb$_4$RhAl in zero field (and in 5 kOe for χ'), measured with different frequencies. Inset shows χ' plots in an expanded form around 28 K to show the frequency dependence. χ" curves in 5 kOe are featureless and hence are not shown.

Figure 10:
(a) Electrical resistivity as a function of temperature (1.8-100 K) for Tb$_4$RhAl in the presence of different external magnetic fields. Inset shows the zero-field data up to 300 K. (b) Magnetoresistance as a function of magnetic field (0 – 130 kOe, virgin curve only) for 1.8, 5, 10, 28.5 and 40 K. (c) MR versus $H$ behavior at 1.8 K for a variation of the field 0 to 130 kOe to -130 kOe to 130 kOe to 0 and the arrows mark the way the field is varied.

Figure 11:
Isothermal entropy change as a function of temperature for various final fields with initial zero external field obtained from isothermal magnetization data, measured at close intervals of temperature, as well as from the heat-capacity data measured in the presence of external fields for (a) Gd$_4$RhAl, and (b) Tb$_4$RhAl. The lines through the data points serve as guides to the eyes.

Figure 12:
The plot of isothermal entropy change for $\Delta H = 50$ kOe as a function of $H$ at different temperatures in the vicinity of respective magnetic transition temperatures for (a) Gd$_4$RhAl and (b) Tb$_4$RhAl. The continuous lines through the data points are fits to the power law, $H^n$. The exponent ($n$) values are also indicated.

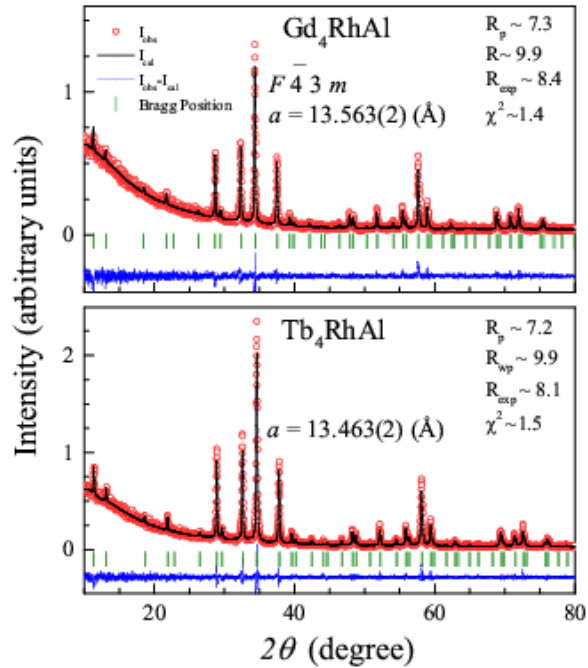

Fig1

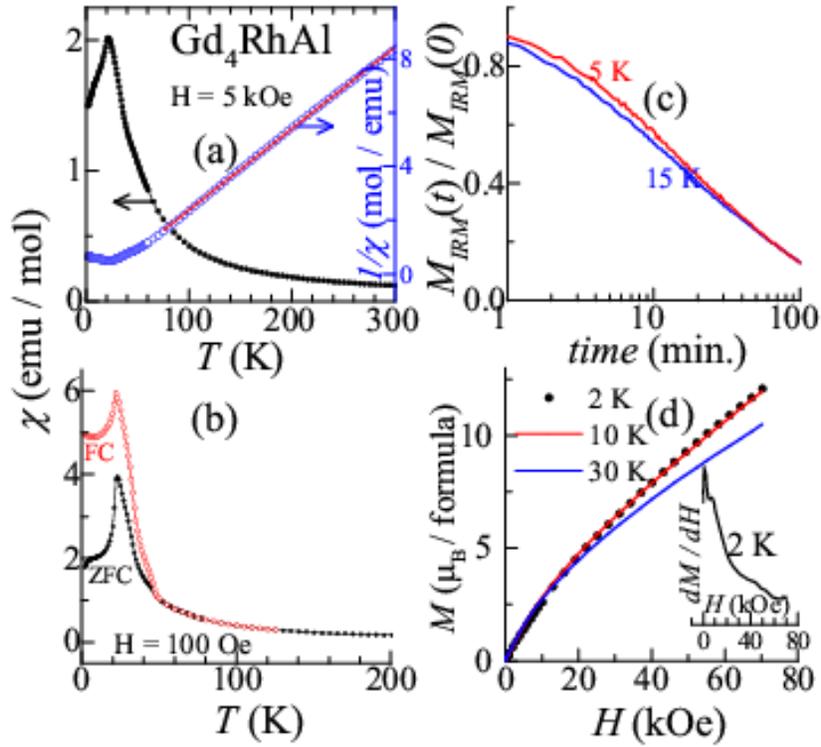

Fig2

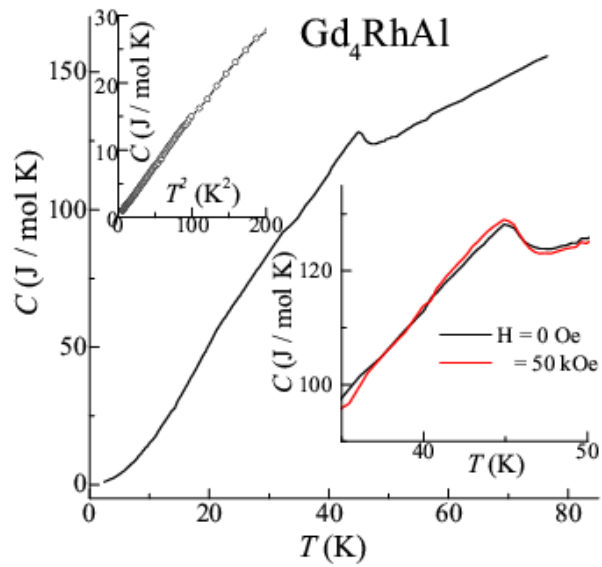

Fig3

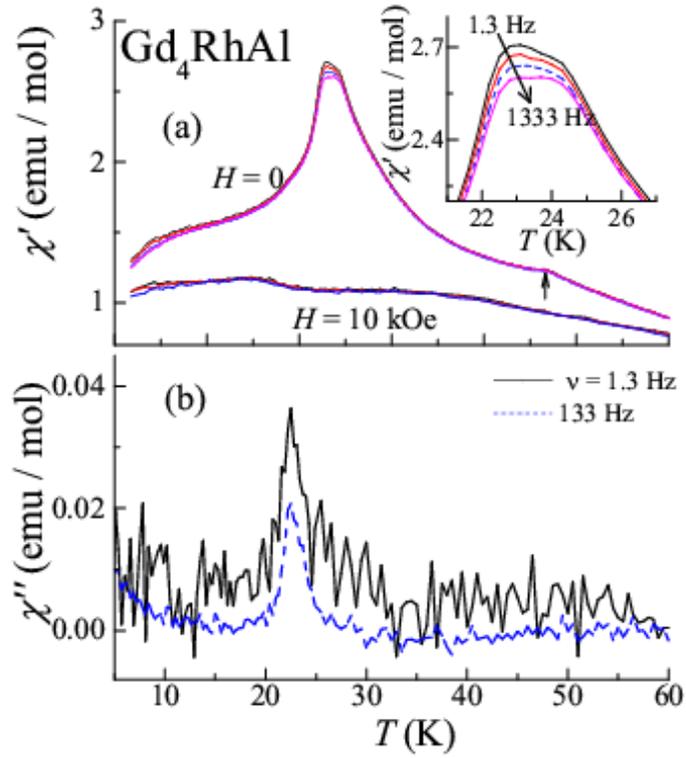

Fig4

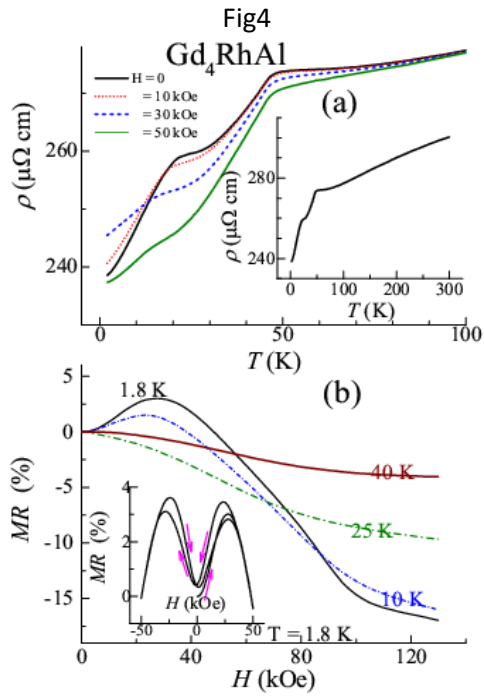

Fig5

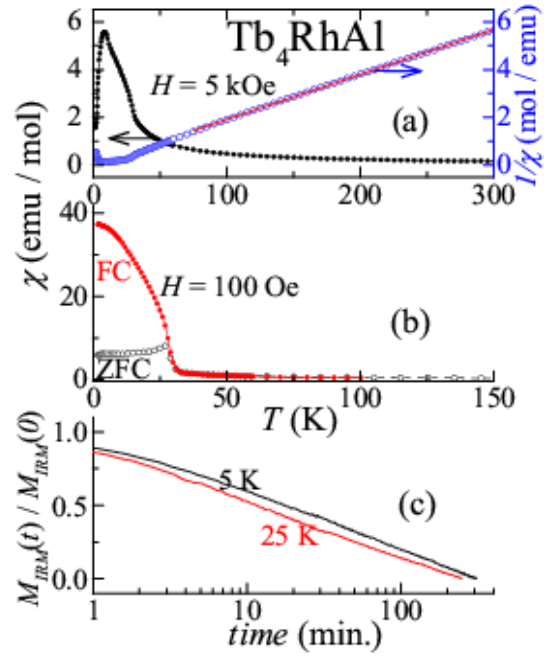

Fig6

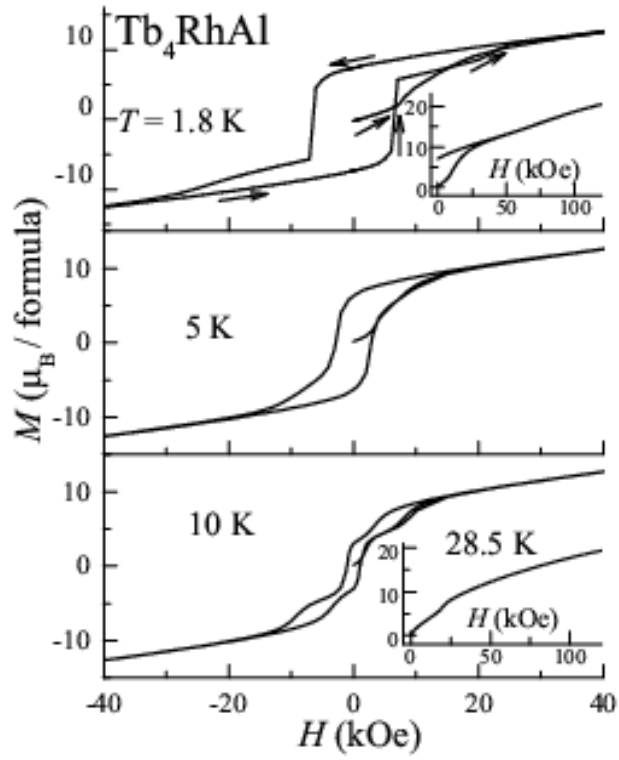

Fig7

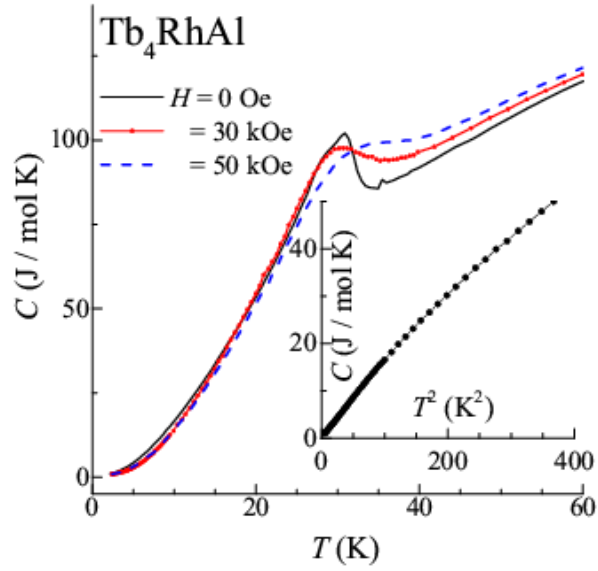

Fig8

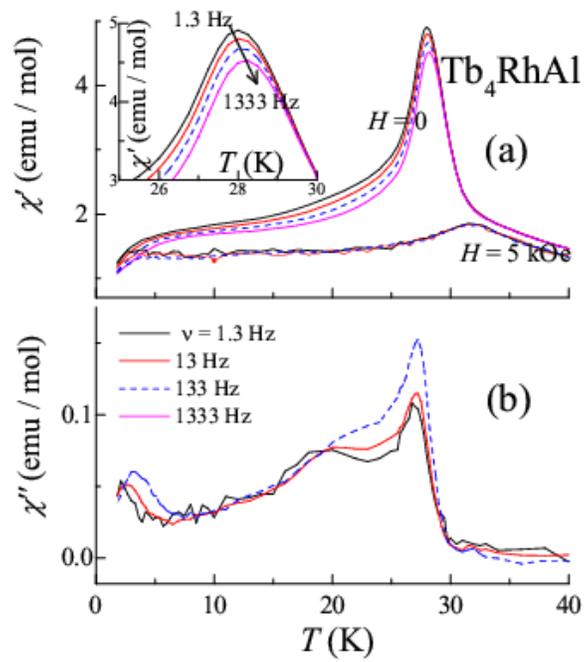

Fig9

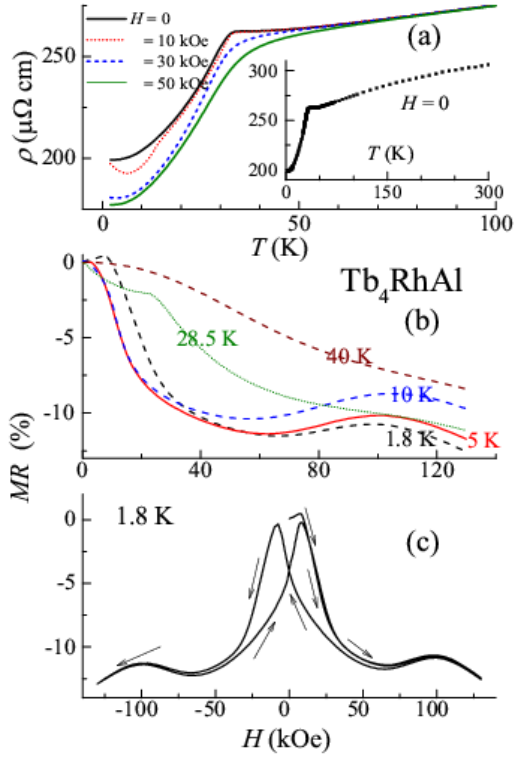

Fig10

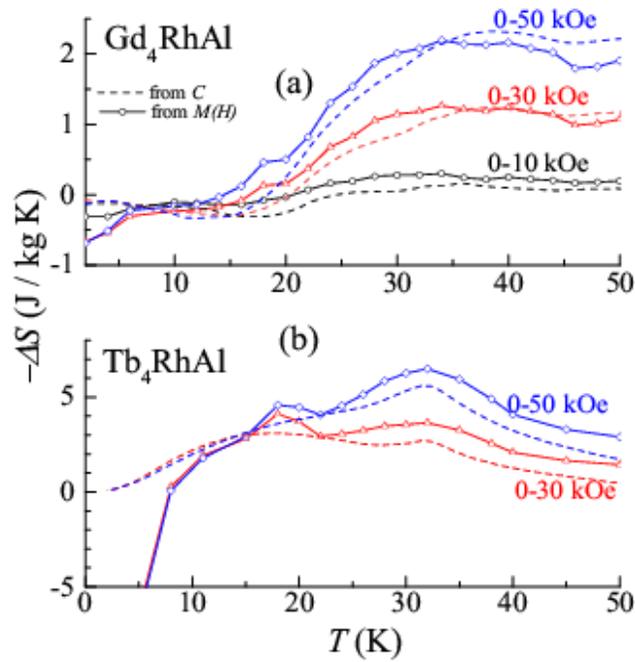

Fig11

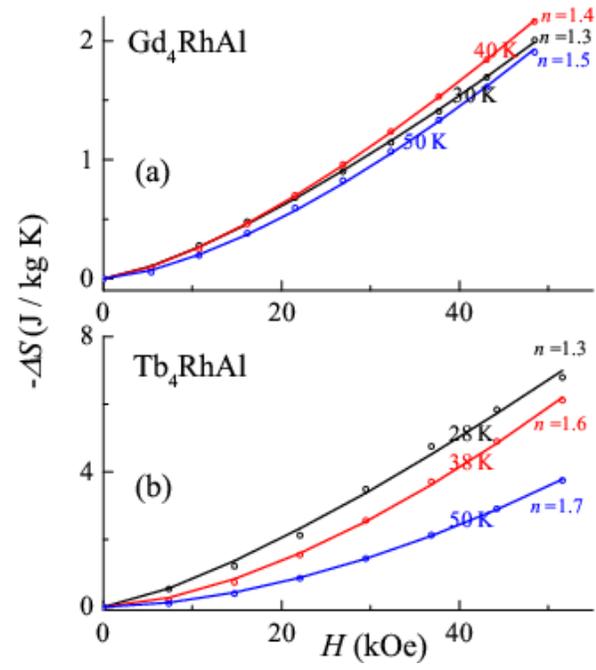
Fig12